\documentclass[aps,prl,twocolumn,10pt,showpacs,superscriptaddress,footinbib]{revtex4-1}
\usepackage{amsmath,amssymb}
\usepackage{graphicx,color,colortbl,overpic}
\usepackage{rotating,array,tabularx,booktabs}
\usepackage[squaren]{SIunits}
\usepackage[normalem]{ulem}
\newcolumntype{Y}{>{\centering\arraybackslash}X}

\newcommand{\dif}{\mathrm{d}}%
\newcommand{\tdif}[2]{\frac{\dif#1}{\dif#2}}%
\newcommand{\ie}{i.\,e.}%

\graphicspath{{Bilder/}}

\begin{document}
\title{Brownian motion and the hydrodynamic friction tensor\\for colloidal particles of complex shape}

\author{Daniela J. Kraft}
\email[Corresponding author: ]{kraft@physics.leidenuniv.nl}
\affiliation{Center for Soft Matter Research, Department of Physics, New York University, New York, NY 10003}
\author{Raphael Wittkowski}
\affiliation{SUPA, School of Physics and Astronomy, University of Edinburgh, Edinburgh, EH9 3JZ, United Kingdom}
\author{Borge ten Hagen}
\affiliation{Institut f{\"u}r Theoretische Physik II, Weiche Materie, Heinrich-Heine-Universit{\"a}t D{\"u}sseldorf, D-40225 D{\"u}sseldorf, Germany}
\author{Kazem V. Edmond}
\affiliation{Center for Soft Matter Research, Department of Physics, New York University, New York, NY 10003}
\author{David J. Pine}
\affiliation{Center for Soft Matter Research, Department of Physics, New York University, New York, NY 10003}
\author{Hartmut L{\"o}wen}
\affiliation{Institut f{\"u}r Theoretische Physik II, Weiche Materie, Heinrich-Heine-Universit{\"a}t D{\"u}sseldorf, D-40225 D{\"u}sseldorf, Germany}

\date{\today}

\begin{abstract}
We synthesize colloidal particles with various anisotropic shapes and track their orientationally resolved Brownian trajectories using confocal microscopy. An analysis of appropriate short-time correlation functions provides direct access to the hydrodynamic friction tensor of the particles revealing nontrivial couplings between the translational and rotational degrees of freedom. The results are consistent with calculations of the hydrodynamic friction tensor in the low-Reynolds-number regime for the experimentally determined particle shapes.
\end{abstract}


\maketitle


%
\begin{figure*}[ht]
\begin{center}
\includegraphics[width=\linewidth]{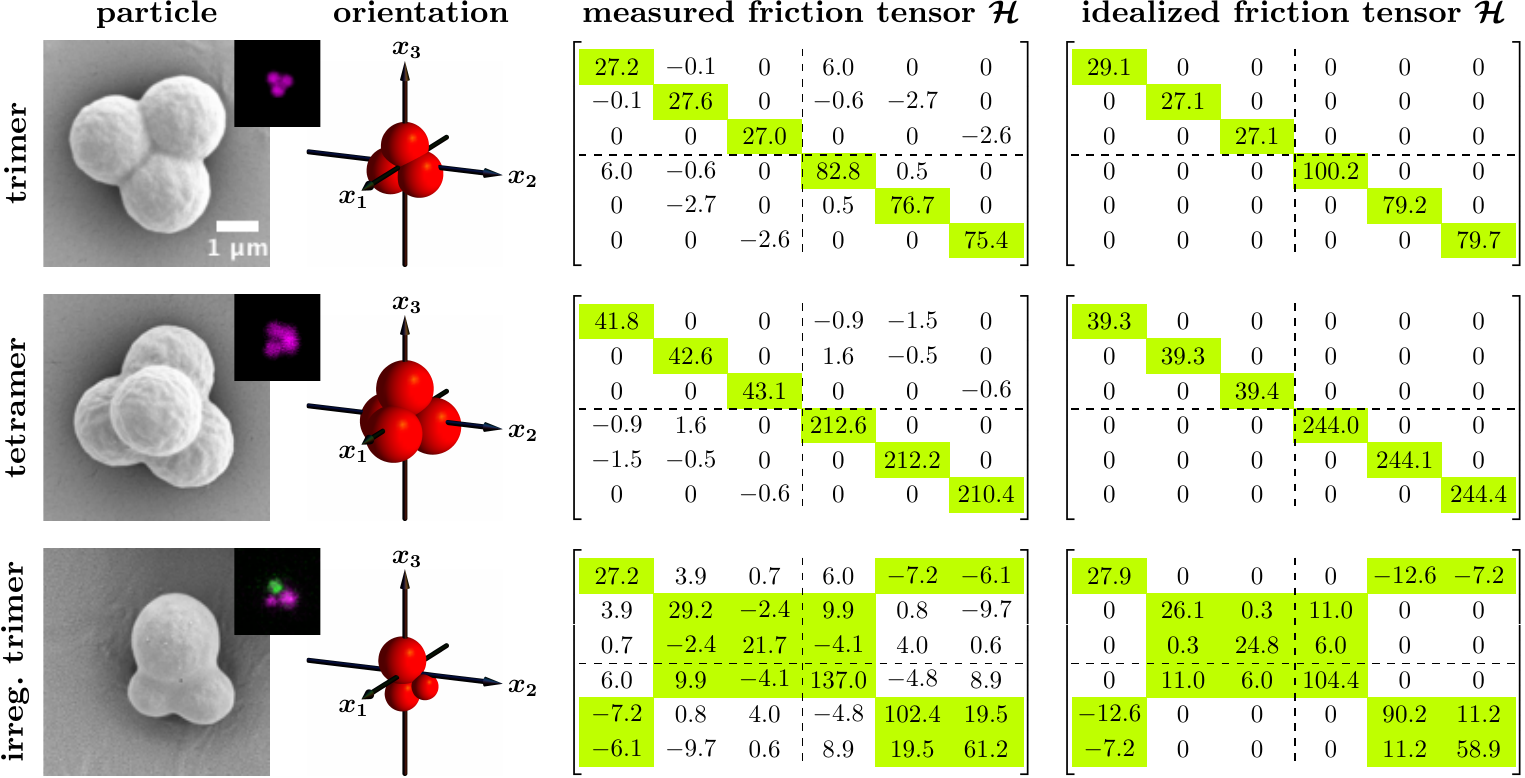}
\end{center}
\caption{\label{fig:Particles}SEM images (gray) and images obtained by confocal microscopy (colored insets) of three different colloidal particles (regular trimer, regular tetramer, and irregular trimer) and the corresponding hydrodynamic friction tensors $\mathcal{H}$ determined in experiments and predicted for idealized particle shapes by hydrodynamic calculations using \texttt{Hydrosub}. The tensors shown are given in dimensions $[\mathcal{H}_{i,j=1,2,3}]=\micro\meter$, $[\mathcal{H}\begin{subarray}{l}i=1,2,3\\j=4,5,6\end{subarray}]=[\mathcal{H}\begin{subarray}{l}i=4,5,6\\j=1,2,3\end{subarray}]=\micro \squaren \meter$, and $[\mathcal{H}_{i,j=4,5,6}]=\micro\meter\cubed$. 
The four $3\times 3$ submatrices of $\mathcal{H}$ contain the translational friction coefficients (upper left), the rotational friction coefficients (lower right), and the translational-rotational friction coefficients for the coupling of translational and rotational motion (upper right and lower left). 
Since $\mathcal{H}$ depends on the center-of-mass position and orientation of the particles, the coordinate systems used are illustrated.
For the diagonal elements of $\mathcal{H}$, the statistical error of the experimental data is $3\%$ for the regular trimer and tetramer and $10\%$ for the irregular trimer. The absolute statistical error of the off-diagonal elements is $1$ and $5$, respectively, in the given units.}
\end{figure*}
\begin{figure*}[ht]
\begin{center}
\includegraphics[width=1.0 \linewidth]{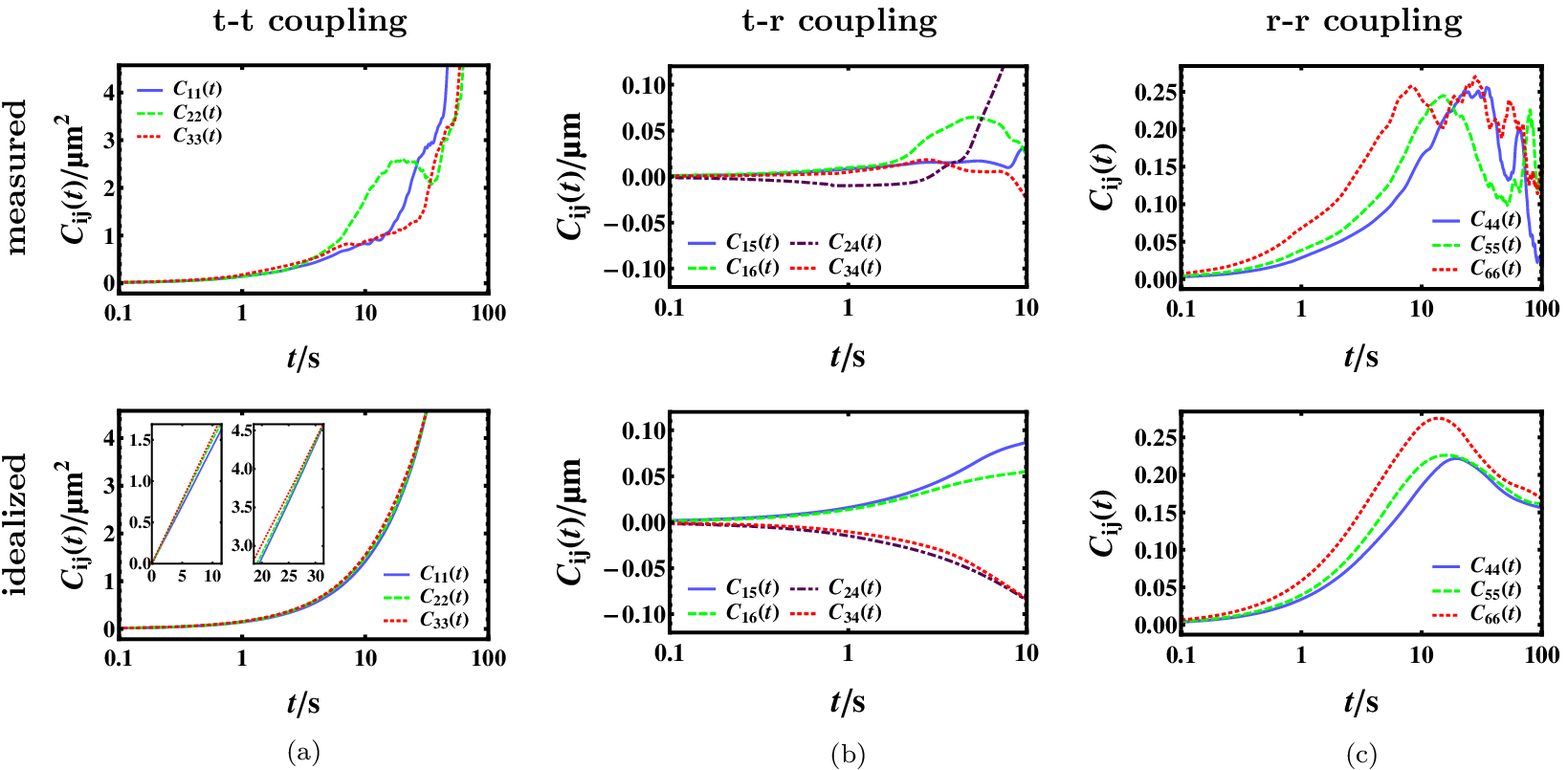}
\end{center}
\caption{\label{fig:Cij}Time evolution of ten representative correlation functions $C_{ij}(t)$ for the irregular trimer determined in experiments (upper row) and predicted from a simulation (lower row) based on the idealized hydrodynamic friction tensor (see Fig.\ \ref{fig:Particles}). The insets in the lower left plot show the same quantities on linear scales.}
\end{figure*}
%


Spherical colloidal particles have served as a model system for investigating Brownian motion since the pioneering studies of A.\ Einstein \cite{Einstein1905c} and J.\ Perrin \cite{Perrin1909}. Such particles are characterized by a translational diffusion coefficient that is linked to the Stokes friction coefficient through the well-known Stokes-Einstein relation \cite{Einstein1905c}. Particles encountered in nature and industry usually have complex non-spherical shapes, and describing their Brownian motion raises fundamental questions about how translational and rotational diffusion are coupled. However, in most studies, translational and rotational diffusion are considered separately, which is valid only for certain highly symmetrical particle shapes.  
 
Recently, model colloids with well-characterized but complex shapes have become available \cite{ManoharanEP2003,GlotzerS2007,*Kraft2009JACS,*Kraft2009SoftMatter}, which permits the quantitative study of the hydrodynamic coupling between translational and rotational diffusion for nontrivial particle shapes for the first time.
 
In general, the dynamics of a colloidal particle suspended in a liquid is described by a Langevin equation that equates the Stokes friction forces and torques with random thermal forces and torques on a particle. 
For an arbitrary colloidal particle suspended in a liquid, the friction forces and torques are described by a symmetric second-rank hydrodynamic friction tensor $\mathcal{H}$ \cite{Brenner1967,HappelB1991}, which includes off-diagonal terms coupling the three translational and three rotational degrees of freedom.
In all, $\mathcal{H}$ has $21$ independent elements. 
For spherical particles, $\mathcal{H}$ is diagonal, with two distinct entries corresponding to the inverse translational and rotational friction coefficients \cite{Naegele1996,*DegiorgioPJ1995}.
For rod-like particles, both the translational and rotational entries involve two different coefficients, corresponding to parallel and perpendicular particle orientation, but $\mathcal{H}$ remains diagonal, meaning that translation and rotation remain decoupled \cite{Dhont1996,*HanANZLY2006}.
 
For a general biaxial particle, $\mathcal{H}$ involves nonzero off-diagonal elements that couple translational and rotational motion.
The corresponding Langevin equation involves intricate multiplicative noise terms due to this coupling, which makes a description of the Brownian dynamics much more difficult. 
Although a first theoretical treatment dates back to F.\ Perrin \cite{Perrin1934,*Perrin1936}, it was not reconsidered until much later, and only  by a few authors \cite{Brenner1967,FernandesdlT2002,*MakinoD2004,*WittkowskiL2012} who never explicitly applied it to experiments for biaxial non-orthotropic particle shapes \footnote{Notice that there is related earlier work that considers only the translational part of the friction tensor \cite{HubbardD1993,*BoukariNSS2004} or just the viscosity \cite{BiceranoDB1999}.}.

In this article, we report experimental measurements and theoretical calculations of the hydrodynamic friction tensor for various anisotropic colloidal particles, including a general irregular biaxial shape with three fused spheres of different diameters.
The particle shape and size are determined by confocal as well as scanning electron microscopy (SEM).
We track the Brownian trajectories of these anisotropic colloidal particles with full orientational resolution in real space by confocal microscopy. This 3D real-space technique allows for tracking the motion of arbitrarily complex colloidal particles, even in crowded environments. Based on the generalized Stokes-Einstein relation, we then propose a theoretical framework to extract all independent hydrodynamic friction coefficients from the short-time limit of appropriate correlation functions. Our results are consistent with low-Reynolds-number hydrodynamic calculations of the friction tensor assuming stick boundary conditions of the solvent at the particle surface, where the experimentally determined particle shapes are taken as an input. 
Since the full orientational resolution of the individual particle trajectories reveals the couplings between different degrees of freedom of Brownian motion, the information obtained by our analysis is much more basic and detailed than averaged quantities derived from dynamic light scattering \cite{HoffmannWHW2009} or sedimentation \cite{ZahnLM1994} experiments of biaxial colloidal particles. 
Our method can be used to analyze the Brownian dynamics of any rigid irregularly-shaped colloidal particles.

For our experiments, we have prepared anisotropic colloidal particles from fluorescently labeled, cross-linked PMMA (poly(methyl methacrylate)) spheres \cite{Elsesser:2011fp} using an emulsion-evaporation method \cite{ManoharanEP2003}. The resulting cluster shapes are uniquely set by minimization of the second moment of the mass distribution \cite{ManoharanEP2003} as confirmed by SEM. We have specifically chosen regular trimers and tetramers as well as an irregular trimer shown in Fig.\ \ref{fig:Particles} for their different symmetry properties. We idealized the particle cluster shapes as a composition of fused spheres and measured only the radii and relative distances between the spheres.

For the regular clusters, RITC (rhodamine B isothiocyanate) labeled PMMA spheres $\unit{2.1\pm 0.1}{\micro\meter}$ (trimer) and $\unit{2.4\pm0.1}{\micro\meter}$ (tetramer) in diameter are employed. 
For the irregular trimer, $\unit{2.1\pm0.1}{\micro\meter}$ and $\unit{1.3\pm0.1}{\micro\meter}$ RITC dyed spheres are combined with $\unit{1.7\pm0.1}{\micro\meter}$ spheres labeled with NBD-MAEM (4-methylaminoethyl methacrylate-7-nitrobenzo-2-oxa-1,3-diazol), which allows us to easily distinguish the different spheres with a confocal microscope. All sphere diameters are measured by static light scattering (SLS). 
The clusters are dispersed in a TBAB (tetra butyl ammonium bromide) saturated $\sim\!78/22$ (weight/weight) cyclohexyl bromide/\textit{cis}-decalin (CHB/decalin) mixture, which nearly matches the particles' density and index of refraction, and has a dynamic (shear) viscosity $\eta=\unit{2.22}{\milli \pascalsecond}$. The dispersion is then put into rectangular glass capillaries (Vitrotubes, $\unit{100}{\micro\meter}\times \unit{5}{\milli\meter} \times \unit{50}{\milli\meter}$) and the ends are sealed with optical adhesive (Norland \#81).

The three-dimensional motion of the anisotropic particles is observed using a Leica TCS SP5 confocal microscope equipped with an argon laser ($\lambda_{1}=\unit{488}{\nano\meter}$ and $\lambda_{2}=\unit{543}{\nano\meter}$) and an oil-immersion objective (Leica, $63\times$, $1.4$ NA). The imaging speed is typically $70$ stacks in $z$-direction per approximately $\unit{0.8}{\second}$. All experiments are conducted at room temperate $T=\unit{294}{\kelvin}$. The positions of the individual spheres of each cluster are tracked using IDL routines (see Ref.\ \cite{Hunter:2011ew}). From these sphere positions we calculate the center-of-mass positions, orientations, and bond lengths of the clusters. For the regular trimer the bond length is $\unit{1.5}{\micro\meter}$, for the regular tetramer the centers of any two spheres are separated by $\unit{2.3}{\micro\meter}$, and for the irregular trimer the bond lengths are $\unit{2.2}{\micro\meter}$ between the big and the medium sphere as well as between the big and the small sphere, and $\unit{1.7}{\micro\meter}$ between the medium and the small sphere.

Apart from temperature $T$ and solvent viscosity $\eta$, the Brownian dynamics of a single rigid colloidal particle depends only on its shape and size, which enter in the $6\!\times\!6$-dimensional symmetric hydrodynamic friction tensor $\mathcal{H}$ \cite{Brenner1967,HappelB1991}. 
The latter relates the translational velocity $\vec{v}$ and the angular velocity $\vec{\omega}$ of the particle 
to the hydrodynamic drag force $\vec{F}$ and torque $\vec{T}$ that the particle experiences in the viscous solvent: 
$\vec{K}=-\eta\,\mathcal{H}\,\vec{\mathfrak{v}}$ with $\vec{K}=(\vec{F},\vec{T})$ and $\vec{\mathfrak{v}}=(\vec{v},\vec{\omega})$. 

There are two possibilities for determining $\mathcal{H}$ for a given particle. 
It can either be obtained from its shape and size by a hydrodynamic calculation that involves solving the Stokes equation with stick boundary conditions for the solvent at the particle surface \cite{HappelB1991}, or it can be extracted from appropriate equilibrium short-time correlation functions.
We have used the software \texttt{Hydrosub} \cite{delaTorreB1981,*CarrascoT1999,*delaTorreC2002} to follow the first route, where we used the experimentally determined particle shape, idealized by fused spheres, as input \cite{AbadeCEJNW2010}. 
For a trimer and a tetramer of equal spheres as well as for an irregular trimer, results are shown in Fig.\ \ref{fig:Particles} \footnote{Clearly, our method is also applicable to dimers, where the tensor $\mathcal{H}$ is diagonal. We obtain agreement with the theoretical calculations within a few percent.}. 
For convenience, we have chosen the coordinate systems in such a way that the center of mass of a particle coincides with the origin of coordinates and the particle's planes of symmetry coincide with the coordinate planes, whenever this is possible.
This choice of particle-fixed coordinate systems leads to a particularly simple structure of the hydrodynamic friction tensor with many vanishing non-diagonal elements \cite{HappelB1991}. The remaining non-vanishing elements are highlighted in Fig.\ \ref{fig:Particles}.

The second route to access $\mathcal{H}$ is to measure the trajectory of the Brownian particle with full orientational resolution, \ie, the combined knowledge of the center-of-mass position $\vec{x}(t)$ and the three mutually perpendicular normalized orientation vectors $\hat{u}_{i}(t)$ with $i=1,2,3$ in dependence of time $t$.
The key idea is now to consider a set of appropriate dynamical cross-correlation functions 
\begin{equation}
C_{ij}(t)=\langle X_{i}(t)X_{j}(t)\rangle
\end{equation}
with $i,j\in\{1,\dotsc,6\}$, where $\langle\,\cdot\,\rangle$ denotes a noise average and the six-dimensional positional-orientational displacement vector $\vec{X}(t)=(\Delta\vec{x}(t),\Delta\hat{u}(t))$ is defined by $\Delta\vec{x}(t)=\vec{x}(t)-\vec{x}(0)$ and $\Delta\hat{u}(t)=\frac{1}{2}\sum^{3}_{i=1}\hat{u}_{i}(0)\times\hat{u}_{i}(t)$, where the latter is the appropriate expression for orientational displacements. 
The short-time limit of this set of cross-correlation functions gives access to the hydrodynamic friction tensor $\mathcal{H}$ via
\begin{equation}
\mathcal{D}=\frac{1}{2}\lim_{t\to 0}\tdif{C(t)}{t} \;, 
\qquad\mathcal{H}=\frac{k_{\mathrm{B}}T}{\eta}\,\mathcal{D}^{-1} \;,
\label{eq:H}
\end{equation}
where $\mathcal{D}$ denotes the (generalized) diffusion tensor and $k_{\mathrm{B}}$ Boltzmann's constant. 
A larger value for an element of $\mathcal{H}$ therefore means a higher hydrodynamic friction and thus a slower diffusion.  
From this second route, based on the experimentally determined trajectories, we obtain the results presented in Fig.\ \ref{fig:Particles}. 
The experimental results for $\mathcal{H}$ are in good agreement with our hydrodynamic calculations; deviations are due to an idealization of the particle shape in the hydrodynamic calculations and due to the statistical error originating from the limited length of the measured trajectories. 

In reality, the particles are not compositions of perfect spheres, but have rough surfaces and deformations near the overlap-areas of the spheres. Additionally, the spheres that make up the clusters have a polydispersity of about $5\%$. While uncertainties in the size of the particles only lead to small deviations in the translational-translational elements $\mathcal{H}_{i,j=1,2,3}\propto l$ with the length scale $l$, these deviations are of greater relevance for the translational-rotational coupling elements $\mathcal{H}\begin{subarray}{l}i=1,2,3\\j=4,5,6\end{subarray}\propto l^{2}$ and lead to large deviations in the rotational-rotational elements $\mathcal{H}_{i,j=4,5,6}\propto l^{3}$. 
Non-diagonal tensor elements, which should vanish by symmetry considerations (not highlighted in Fig.\ \ref{fig:Particles}), indeed have nonzero values in the experimentally determined friction tensor due to both irregularities in the actual particle shape and the statistical error because of the finite length of the trajectories.

We finally address the full time dependence of the basic cross-correlation functions $C_{ij}(t)$. 
Results for $C_{ij}(t)$ obtained from our measured trajectories of the irregular trimer particle are presented in Fig.\ \ref{fig:Cij}.
Figure \ref{fig:Cij}(a) shows the translational mean-square displacements for the different Cartesian components, which increase linearly for short times with the slope governed by the anisotropic short-time diffusion coefficients. At long times, there is  another linear function in time, which is the same for all coordinates \cite{CichockiEJW2012}, since it is governed by the orientationally-averaged long-time diffusion coefficient [see insets in the lower plot in Fig.\ \ref{fig:Cij}(a)]. 
Several cross-correlations between translational and rotational displacements are shown in Fig.\ \ref{fig:Cij}(b). 
The absolute values of these nontrivial correlations are initially increasing with time, but decorrelate for longer times.  
Finally, the rotational-rotational correlations shown in Fig.\ \ref{fig:Cij}(c) are clearly positive and build up continuously as functions of time until they decay again and approach the same constant value for very long times.
Here, we restrict the presentation of correlation functions to this irregular particle, since it is the only particle with nonzero translational-rotational coupling elements and thus provides the most nontrivial dynamics. 

An idealized time evolution of the correlation functions $C_{ij}(t)$ can be obtained from a given hydrodynamic friction tensor $\mathcal{H}$. With this tensor as an input, the Brownian motion of a colloidal particle can be simulated by solving its Langevin equation (see Ref.\ \cite{WittkowskiL2012}) numerically using a stochastic integrator of strong order 1.5 \cite{KloedenP2006}. Our results for the predicted correlation functions are presented in Fig.\ \ref{fig:Cij} as well. 
A comparison with the experimental results for $C_{ij}(t)$ reveals again good agreement with deviations resulting from choosing the idealized instead of the experimentally determined hydrodynamic friction tensor and from the statistical error.
Note that the statistical error is obvious for the experimentally determined correlation functions, while it is extremely small for the simulated correlation functions, where trajectories with $10^{6}$ time steps have been calculated. 

In this work, we track the Brownian dynamics of individual colloidal particles with various anisotropic shapes and extract the hydrodynamic friction tensor from an analysis of appropriate short-time correlation functions. The framework of our
analysis can in principle be applied to any rigid particle with an arbitrary shape and therefore to a broad range of relevant suspensions. Using confocal microscopy we obtain real-space 3D measurements of any complex particles, even in crowded environments. 

Future work should address the effect of aligning external fields such as gravity or magnetic fields \cite{vanderBeekDWVL2008} that gain major importance in the context of directed self-assembly \cite{GrzelczakVFLM2010}. Moreover, non-dilute colloidal suspensions would be interesting, where direct particle-particle interactions \cite{Yethiraj2007} and solvent-flow mediated hydrodynamic interactions \cite{vandeVen_book} will lead to even more intricate translational-rotational couplings. This would provide a basis to understand the rheological properties of concentrated dispersions of 
irregularly shaped colloidal particles like clay \cite{KegelL2011} and asphalt \cite{EvdokimovEE2001}.

\acknowledgments{We thank Tom Lubensky, HyunJoo Park, Gary Hunter, Andrew Hollingsworth, and Marco Heinen for helpful discussions.
The theoretical work was supported by the German Research Foundation (DFG) within SFB TR6 (project D3) and by the European Research Council (ERC Advanced grant INTERCOCOS, project number 267499). 
The experimental work was supported by the US National Science Foundation (CBET-1236378 and DMR-1105455).
D.\ J.\ K.\ gratefully acknowledges financial support through a Rubicon Grant (680-50-1019) by the Netherlands Organization for Scientific Research (NWO).
R.\ W.\ gratefully acknowledges financial support through a Postdoctoral Research Fellowship (WI 4170/1-1) from the DFG.}

\bibliography{refs}
\end{document}